Comment to: "Electric field vector mapping of guided ionization waves at atmospheric pressure" (ArXiv :1709.03109v1)

Eric Robert, Jean Michel Pouvesle, Xavier Damany, Sébastien Dozias and Thibault Darny
GREMI, UMR 7344 CNRS / Université d'Orléans, Orléans, 45067, France

Abstract
Recently, Iséni (ArXiv :1709.03109v1) reported measurements and analysis of electric field (EF) strengths adjacent to propagating ionization waves (IWs) in non-thermal atmospheric pressure He plasma jets. We demonstrate that these measurements are in error due to an improperly aligned electric field probe.

Comment

Iséni [0] recently reported measurements of guided ionization waves (IWs) generated by an atmospheric pressure plasma jet (APPJ) device in a "plasma gun" configuration. He reported measurements of the electric fields (EFs) driving the IW, with both spatial and temporal resolution, along the direction of IW propagation. We demonstrate that these measurements are inaccurate, apparently due to a misaligned electric field probe.

Furthermore, we wish to make clear that [0] was written solely by Dr. Sylvain Iséni. The inclusion of four of the present authors' names in the acknowledgement section of [0] implies that the present authors support or are in agreement with the measurements and analyses contained in [0]. This is false. Indeed, as noted above and detailed below, we believe [0] contains fundamental flaws in both experimental execution and subsequent analysis and we therefore are in strong disagreement with the reported results in [0].

In previous experiments, and as documented in multiple publications, we describe measurements of the two orthogonal components of the EF in the exact experimental configuration as that reported in figure 1 of [0] with the Kapteos EF probe set 75 mm downstream from the high voltage inner electrode tip (cf. figure 2 of [0]). This measurement of EF components and the orientation of the capillary discharge are shown in figure 1.

Figure 1 shows EF components with typical temporal evolutions and EF amplitudes in perfect agreement with previously measured and simulated data [1-4]. It must be mentioned that such EF signals are measured a few mm radially apart from the capillary and plume axis, all along the propagation path of the ionization wave (IW) launched from the HV electrode down to the tip of the plasma plume.

The most striking difference between the measurement documented in figure 1 and that published by S. Iséni in figure 2 (continuous line) of [0] is observed for the longitudinal ($E_x$) EF temporal evolution.

We assert that the reason for the disagreement between our results and the erroneous results published in [0] is that the EF probe used in [0] was improperly positioned. An improperly positioned (i.e. tilted) probe with respect to the vertical (longitudinal) and horizontal (radial) axis leads to a complete mixing of the two (longitudinal and radial) EF components. Consequently the analysis of IW features, i.e. the ionization front and the plasma column tagged as the Pulsed Atmospheric-pressure Plasma Stream (PAPS) [5], is quite impossible without a correct measurement of longitudinal and radial EF components.

Figure 2 presents the measurement of the two EF components for the same conditions as for figure 1 but with the EF probe tilted by an angle of 45 degree with respect to the vertical axis. This improper orientation is sketched in the inset of figure 2.

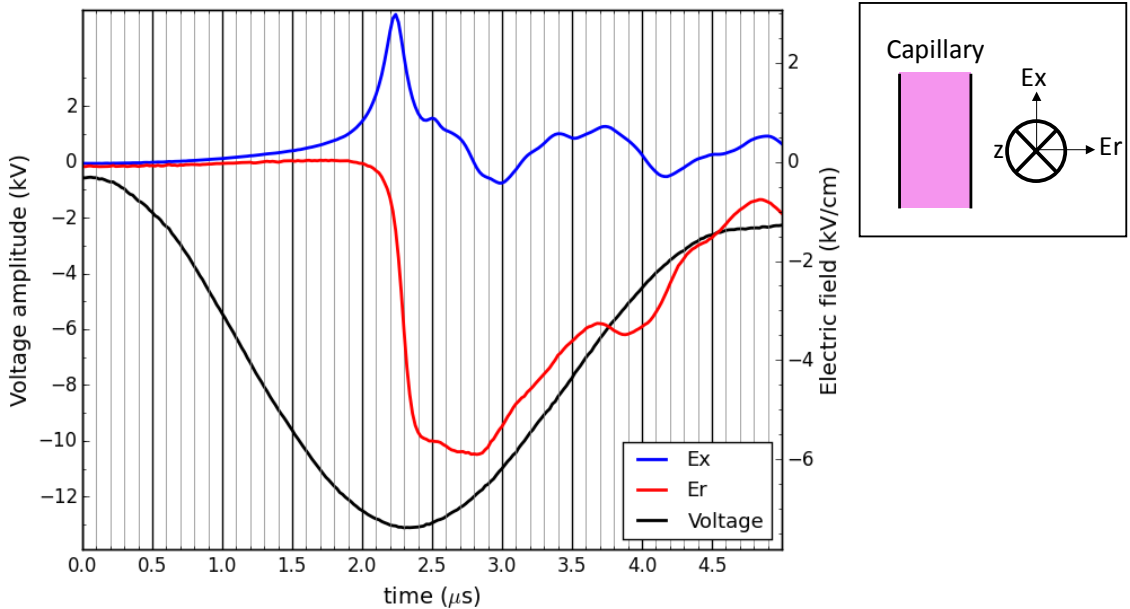

*Figure 1* Applied voltage, longitudinal ($E_x$) and radial ($E_r$) EF temporal evolutions during one voltage pulse. The inset in the upper right is a side view of the cylindrical capillary discharge and a schematic of the EF probe positioning allowing for the two EF components (($E_x$) and ($E_r$)) measurement when the probe is properly aligned. Measurements made 75 mm from the powered inner electrode (cf. figure 2 of [0])

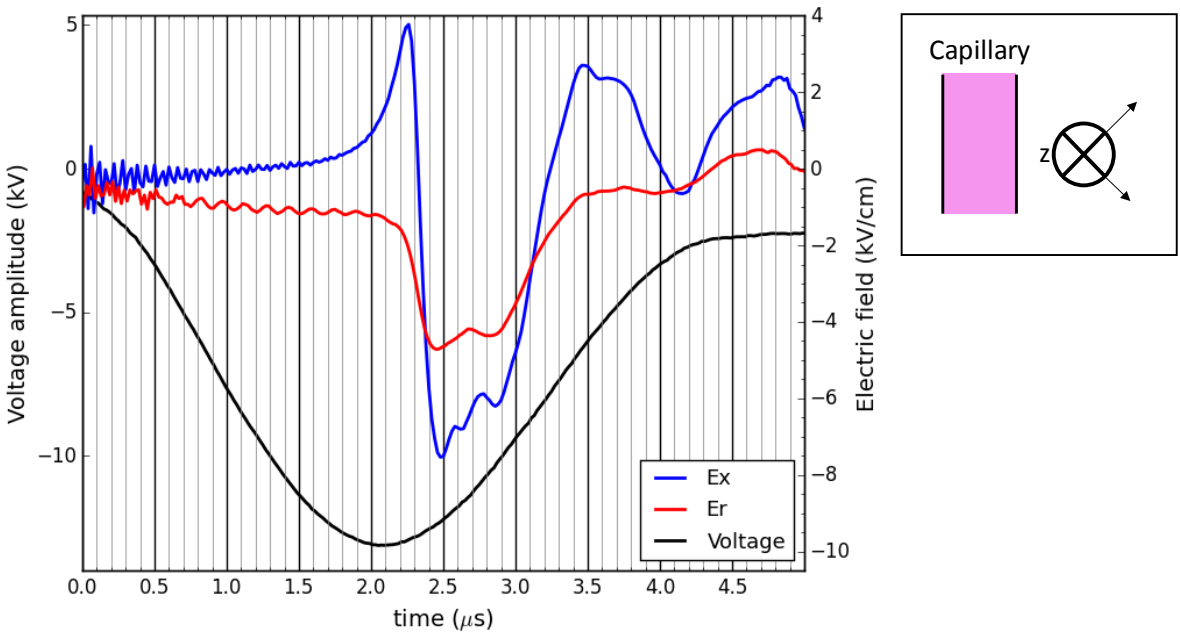

*Figure 2* Applied voltage, longitudinal and radial EF temporal evolutions as shown in figure 1, but with an improper 45° degree tilt of the EF detector, as sketched in the inset.

We note that with the improperly tilted probe configuration as illustrated in figure 2, the measured EF components' temporal evolution now perfectly match those documented in figure 2 of [0]. This not only confirms our critical interpretation but definitely demonstrates that data collection and analysis inferred from data in [0] are incorrect. All reported results in [0] such as "intriguing configuration of EF lines", are only artifacts due to improper orientation of the EF probe. This erroneous orientation of the probe leads to incorrect $E_x$ and $E_r$ amplitudes as well as mappings of EF lines.

Finally, the present authors would welcome anyone who wishes to compare their results with the ones we have published here and in the previous literature to visit our laboratory (GREMI, Université d'Orléans, Orléans, France). The EF probe and the plasma jet test bench in our lab will be made available to interested scientists.